\documentclass[12pt, hidelinks]{article}
\usepackage[utf8]{inputenc}
\usepackage{amsmath, amssymb, url, float, graphicx, float}
\usepackage{fullpage}
\usepackage[small,bf]{caption}
\usepackage{cite}

\renewcommand\phi\varphi
\newcommand{\eps}{\varepsilon}

\let\phi\varphi

\newcommand{\ones}{\mathbf 1}
\newcommand{\reals}{{\mbox{\bf R}}}

\newcommand{\cf}{{\it cf.}}
\newcommand{\eg}{{\it e.g.}}
\newcommand{\ie}{{\it i.e.}}

\newcommand{\BEAS}{\begin{eqnarray*}}
\newcommand{\EEAS}{\end{eqnarray*}}
\newcommand{\BEA}{\begin{eqnarray}}
\newcommand{\EEA}{\end{eqnarray}}
\newcommand{\BEQ}{\begin{equation}}
\newcommand{\EEQ}{\end{equation}}
\newcommand{\BIT}{\begin{itemize}}
\newcommand{\EIT}{\end{itemize}}

\title{A Note on Privacy in Constant Function Market Makers}
\author{Guillermo Angeris\\\texttt{\small angeris@stanford.edu} \and
Alex Evans\\\texttt{\small alex@placeholder.vc} \and
Tarun Chitra \\ \texttt{\small tarun@gauntlet.network}}
\date{February 2021}

\begin{document} 
\maketitle 

\begin{abstract}
Constant function market makers (CFMMs) such as Uniswap, Balancer, Curve, and mStable, among many others, make up some of the largest 
decentralized exchanges on Ethereum and other blockchains. Because all transactions are public in current implementations,
a natural next question is if there exist similar decentralized exchanges which are privacy-preserving; \ie, if a transaction's
quantities are hidden from the public view, then an adversary cannot correctly reconstruct the traded quantities from other public
information. In this note, we show that privacy is impossible with the usual implementations of CFMMs under most reasonable
models of an adversary and provide some mitigating strategies.
\end{abstract}

\section*{Introduction}

Decentralized exchanges (DEXs) have experienced rapid growth in liquidity and trading volume over the last year. Much of this growth can 
be attributed to the rise of constant function market makers (CFMMs) that allow for computationally cheap on-chain 
trading~\cite{angerisImprovedPriceOracles2020}. This growth has, in turn, motivated attempts to improve existing mechanisms for 
decentralized exchange. For example, current DEX designs do not support private trading, as the full details of each trade that users 
make can be directly attributed to their on-chain identities. In addition to other challenges, the lack of privacy also makes it easier 
for third parties to front-run a user's trades~\cite{daian2020flash, torres2021frontrunner}.
Another such problem is the ability for attackers to deanonymize agents by doing basic statistical analyses
of public trades performed on DEXs~\cite{goldsmith2020analyzing, chainalysis20202020}.
In contrast, centralized brokers and exchanges preserve user privacy, but agents are required to trust that the exchange won't leak sensitive trade data. A natural question to ask is whether or not popular decentralized exchanges such as Uniswap~\cite{angeris2019analysis, zinsmeisterUniswapV2Core} can be adapted to preserve privacy.

The advent of smart contract systems that utilize zero-knowledge proof systems, such as Zexe~\cite{bowe2020zexe}, suggest that it should be possible to privately execute CFMM transactions.
Indeed, a number of proposed protocols such as SecretSwap~\cite{powers_2021} and Manta~\cite{chumanta}, propose potentially privacy-preserving modifications to Uniswap via the use of either trusted hardware or cryptographic improvements.
However, it has been informally and heuristically noted that `black-box' applications of privacy-preserving technology to Uniswap are unlikely to preserve privacy~\cite{whitehat_2020} as the timing of a trade implicitly leaks identity within Uniswap and other constant function market makers (CFMMs), and can be used to reconstruct the trade.

In this paper, we formalize this intuition and prove that CFMMs are generically unable to preserve privacy under even relatively weak adversaries.
We construct a model where knowledge of a CFMM trading function or `invariant,' such as Uniswap's famous $xy = k$ model, combined with observations of the time-ordering of trades allows an attacker to recover the traded quantities, provided the agent is able to interact
with the CFMM contract in a meaningful way.
One of the main benefits of the convexity of a CFMM is that it makes the arbitrage problem between exchanges easy~\cite{angeris2019analysis}.
Our results illustrate a downside to this: convexity allows an adversary to uniquely recover the traded quantities, assuming that neither
the reserve values nor the traded amounts are known.

\paragraph{Summary.} We give a very basic introduction to CFMMs and describe the attack in~\S\ref{sec:impossibility}. The attack depends
on a uniqueness result for a certain system of nonlinear equations, which we show in~\S\ref{sec:uniqueness} by some basic
tools of convex analysis, for a relatively general family of trading functions. In~\S\ref{sec:extensions} we provide
some basic extensions of the proof to more general CFMMs and slightly weaker attacker models.
Finally, we provide a number of potential mitigation mechanisms that protocol designers can use to improve user privacy in~\S\ref{sec:mitigations}.
While our results are negative, they illustrate that more complex economic mechanisms are needed to preserve privacy than one
might initially assume. We hope these results can be used to guide future private decentralized exchange design.

\section{Impossibility of privacy}\label{sec:impossibility}
We will show that constant function market makers cannot be private in their usual implementations, under most reasonable models
of adversaries. For simplicity, we assume no fees, but discuss an extension to the case with fees in~\S\ref{sec:extensions}.

\subsection{Constant function market makers}

We provide only topically relevant definitions of CFMMs in this short note and refer the reader to~\cite{angerisImprovedPriceOracles2020} for a thorough 
introduction to both the definitions and many of the tools used throughout this paper. The notation here differs slightly for simplicity of the presentation,
but is equivalent to the sufficient condition for path independence in~\cite[\S2.3.2]{angerisImprovedPriceOracles2020} in the case of
two assets. 
\paragraph{Definition.} A \emph{constant function market maker}, or CFMM, is an automated market maker
defined by its \emph{reserve quantities} $R \in \reals_+^n$ and a \emph{trading function} $\psi: \reals_+^n\to \reals$. 
The behavior of CFMMs is very simple: an agent proposes some trade $\Delta \in \reals^n$ where $\Delta_i$ is a positive quantity if $\Delta_i$ of coin $i$ is given to the CFMM while it denotes a negative quantity if it is taken.
The CFMM then checks if the trade satisfies
\[
\psi(R + \Delta) = \psi(R),
\]
\ie, if the trade function $\psi$, depending on the reserves, does not change in value after the proposed trade. If so, the trade is
accepted and the CFMM pays out $-\Delta$ of the traded asset from its reserves $R$, leading to the reserve values being updated as $R \gets R + \Delta$. If not, the trade is rejected and the CFMM does not change its state. Additionally, trades for which $R + \Delta \not \ge 0$ are always rejected since they cannot be fulfilled with the current reserves.

\paragraph{Assumptions.} We will, in general, assume that the trading function $\psi$ is a 
strictly concave, increasing function, which holds for essentially
all CFMMs barring some special cases such as mStable (or constant sum
market makers). This is true for any CFMM whose reachable set~\cite[\S2.3]{angerisImprovedPriceOracles2020}
is a strictly convex set, for all reserves $R$. For example, in the case of Uniswap,
or constant product markets, $\psi(R) = R_1R_2$, which is neither concave nor convex,
but it can be equivalently written as $\psi(R) = \sqrt{R_1R_2}$, which is strictly concave, increasing
whenever $R > 0$. (The notion of equivalence used here is that of~\cite[\S2.1]{angerisImprovedPriceOracles2020},
which we will not discuss further in this note.) We discuss extensions which include trading functions
that are not strictly concave in~\S\ref{sec:extensions}, while we
may generally assume that the function $\psi$ is nondecreasing
without loss of generality~\cite[\S A.1]{angerisImprovedPriceOracles2020}.

\paragraph{Reported price.} As shown in~\cite[\S2.4]{angerisImprovedPriceOracles2020}, we have that the marginal price, $c \in \reals^n_+$ of a fee-less CFMM with reserves $R$ is given by
\[
\nabla\psi(R) = \lambda c,
\]
where $\lambda \ge 0$ is a nonnegative scalar multiplier.

\subsection{Adversary definition and attack}
We assume a very simple, but very general, model of an adversary. In our case, the adversary, who we will call Eve, attempts to 
discover the quantity traded by an agent, called Alice. In our model, Eve is unable to see the exact quantities Alice used to trade 
with the CFMM, but knows when Alice's transaction took place. Eve's only ability is to interact with the CFMM in a state before
Alice's transaction and after the transaction.

\paragraph{Action space.} 
In our case, we will assume that Eve is able to query the marginal price of the CFMM, at the current reserves, and whether a given trade $\Delta$ is valid. (We assume she has access to at least one nonzero valid trade.)
We will also assume, as is generally the case, that Eve knows when Alice's transaction took place and can query the CFMM in its state before and after the transaction.
This assumption could be broken, \eg, if there exist fully-private protocols in which no agent can know the transaction times of any other agent, but such protocols have not yet made it into production.\footnote{Cryptographic primitives such as Verifiable Delay Functions (VDFs)~\cite{boneh2018verifiable} can provide such transaction randomization, but have yet to be used in production networks, let alone those with high-transaction rates.}
We note that this attacker model, where Eve knows only the transaction time, is relatively different than standard attacker models in
the blockchain setting~\cite{garayBitcoinBackbone, badertscherOuroborosChronos} and might therefore be useful to consider
in their own right.

\paragraph{Attack description.} The attack will make repeated use of the following `atom': Eve is always able to
reconstruct the reserve amounts given (a) the marginal price at the current reserves and (b) a single
nonzero feasible trade, by solving a basic nonlinear system of equations.
Using this, it is then enough to simply compute the reserve amounts before and after Alice's trade to recover the
traded amounts. More explicitly, the sequence of the attack is as follows:

\begin{enumerate}
	\item Eve queries the marginal price of the CFMM at the current reserves, to get some vector $c$ and then queries
	any valid nonzero trade $\Delta \ne 0$.
	\item Using this information and the known functional form of $\psi$, Eve can recover
	the reserves $R \in \reals^n_{++}$ by finding a solution to the following nonlinear system of equations in $R$:
	\begin{equation}\label{eq:nonlinear-system}
	\nabla\psi(R) = \lambda c, \quad \psi(R + \Delta) = \psi(R).
	\end{equation}
	The fact that this system has a unique solution (\ie, the true reserves $R$) is a slightly technical point which we will discuss later in this 
	section.
	\item Letting Alice's trade be $\Delta_a$ the new reserves are $R_a = R + \Delta_a$, which are not known to Eve,
	but the CFMM can now be queried in this new state.
	\item Eve then queries the contract again to get a new marginal price $c'$ and then queries any nonzero trade $\Delta'$. She again solves the corresponding system of equations~\eqref{eq:nonlinear-system} to find the new reserves $R_a$:
	\[
	\nabla\psi(R_a) = \lambda c', \quad \psi(R_a + \Delta') = \psi(R_a),
	\]
	to receive $R_a \in \reals^n_{++}$.
	\item Eve then computes $R_a - R = \Delta_a$ to receive Alice's traded values.
\end{enumerate}
From here, it is clear that Eve can always exactly compute the traded amounts from Alice, even if Eve is only given
access to very basic quantities. The only thing that remains to be shown is the uniqueness of the solution $R$ and $R_a$ to
the system of equations. (Existence is guaranteed since the true reserve values $R$ and $R_a$ satisfy the equations, by definition.)
There are, of course, many ways in which Eve can compute a solution to~\eqref{eq:nonlinear-system} given
her known data $c$ and $\Delta$. For example,
using a Newton-type method will likely yield very good practical results for general $\psi$, but the results
are much simpler in some important special cases.

\paragraph{Reserve discovery in Uniswap.} In the case where $\psi(R)$ is a constant product market maker such
as Uniswap, \ie, when $R \in \reals_{++}^2$ and
\[
\psi(R) = \sqrt{R_1R_2},
\]
then~\eqref{eq:nonlinear-system} reduces to a linear system of
equations in $R_1$ and $R_2$. In particular, we have:
\[
\nabla\psi(R) = \frac{\sqrt{R_1R_2}}{2}\left(\frac{1}{R_1}, \frac{1}{R_2}\right) = \lambda (c_1, c_2),
\]
so
\begin{equation}\label{eq:uniswap-isoprice}
\frac12 \sqrt{\frac{R_2}{R_1}} = \lambda c_1, \qquad \frac12 \sqrt{\frac{R_1}{R_2}} = \lambda c_2.
\end{equation}
Multiplying both sides of each equation gives
\[
\frac14 = \lambda^2c_1c_2,
\]
or that $\lambda = (2\sqrt{c_1c_2})^{-1}$, since $\lambda \ge 0$. Plugging this value back into~\eqref{eq:uniswap-isoprice},
we find that
\[
\frac{R_2}{R_1} = \frac{c_1}{c_2},
\]
or that $c_1R_1 = c_2R_2$. Finally, let $\Delta \in \reals^2$ be any feasible trade, then
\[
\psi(R + \Delta) = \sqrt{(R_1 + \Delta_1)(R_2 + \Delta_2)} = \psi(R) = \sqrt{R_1R_2},
\]
which easily simplifies to
\[
\Delta_2R_1 + \Delta_1 R_2 + \Delta_1\Delta_2 = 0.
\]
We can then easily recover $R_1$ and $R_2$, given $c$ and $\Delta$ by solving the following system of linear equations:
\[
\begin{aligned}	
	c_1R_1 - c_2R_2 &= 0\\
	\Delta_2R_1 - \Delta_1R_2 &= -\Delta_1\Delta_2.
\end{aligned}
\]
This system has a unique solution $(R_1, R_2)$ provided $c_1\Delta_1 \ne c_2\Delta_2$, which can be shown to hold for
all feasible trades $\Delta \ne 0$. (We provide a much more general proof, which includes this as a special
case, in~\S\ref{sec:uniqueness}.)

The fact that Uniswap's reserves can be recovered using only the marginal price $c$ and a nonzero
feasible trade $\Delta$ has at least one simple, direct proof, which does not make use of~\eqref{eq:nonlinear-system},
but we provide this special case as an easily-verifiable example of the more
general attack. In fact, in the special case of constant product markets, an adversary only requires the existence
of any two nonzero, distinct feasible trades in order to correctly reconstruct the reserves at
some given point in time. We encourage the reader to try this specific problem as an exercise,
and show that this method extends generally to other CFMMs in~\S\ref{sec:extensions}.

\subsection{Uniqueness of solution}\label{sec:uniqueness}
We will show that the solution of~\eqref{eq:nonlinear-system} is unique in the case that
$\psi$ is an increasing, nonnegative, strictly concave function that is 1-homogeneous; \ie, when
\[
\psi(kR) = k\psi(R)
\]
for any $k \ge 0$. (This includes
constant product and constant mean markets, such as Uniswap and Balancer, as special cases.)
We suspect that uniqueness of the solution can be shown
in the more general setting where the function is not 1-homogeneous but leave this for future work.

\paragraph{Reserves at fixed price.} We will define the set of reserves consistent with the first constraint as:
\[
Q(c) = \{R > 0\mid \nabla\psi(R) = \lambda c, ~ \text{for some} ~ \lambda \ge 0 \}.
\]
In other words, $Q(c)$ is the set of reserves which are consistent with the marginal price of $c$.

Because $\psi$ is a strictly concave, 1-homogeneous function, we will show that this set is a ray, \ie,
it can be written as
\[
Q(c) = \{k R^0\mid k > 0\},
\]
for any $R^0 \ge 0$ with $\psi(R^0) > 0$, satisfying $\nabla\psi(R^0) = \lambda^0 c$ for some $\lambda^0 \ge 0$. Note
that inclusion, $\{k R^0\mid k > 0\} \subseteq Q(c)$, follows immediately from the fact that $\psi$ is 1-homogeneous
and $R^0 \in Q(c)$. On the other hand, showing that $Q(c) \subseteq \{k R^0\mid k > 0\}$ is slightly tricker.

To do this, start with any $R \in Q(c)$ and consider the $\alpha$-superlevel set of $\psi$, given by:
\[
S(\alpha) = \{R \mid \psi(R) \ge \alpha\},
\]
which is a strictly convex set since $\psi$ is a strictly concave function. Additionally,
we will make use of the fact that, for any $k > 0$,
\[
kS(\alpha) = S(k\alpha),
\]
by the homogeneity of $\psi$. (Here $kS(\alpha)$ denotes elementwise set multiplication.)

Given this definition, we know that $c$ is a supporting hyperplane
of the set $S(\psi(R^0))$ at the point $R^0$ and of the set $S(\psi(R))$ at the point $R$. (This follows immediately
from the first-order conditions for convexity applied to the function $\psi$ along with the definition of $Q(c)$.)
Now, because $\psi(R^0) > 0$ and $\psi(R) > 0$, there exists some $k > 0$ such that
$k\psi(R^0) = \psi(R)$. Additionally, we have, by homogeneity,
\[
kS(\psi(R^0)) = S(k\psi(R^0)) = S(\psi(kR^0)) = S(\psi(R)).
\]
But, since $c$ is a supporting hyperplane for $S(\psi(R^0))$ at $R^0$, it is a supporting hyperplane of $kS(\psi(R^0))$ and therefore
of $S(\psi(kR^0)) = S(\psi(R))$ at $kR^0$ and $R$. By strict convexity, every supporting hyperplane of a set will map to a unique
point on the boundary, so we must have that, in fact, $kR^0 = R$, so $R$ lies on the ray generated by $R^0$, as required.

\paragraph{Reserves consistent with a trade.} Now we have to show that the intersection between the set $Q(c)$ and the set
\[
U(\Delta) = \{R > 0 \mid \psi(R + \Delta) = \psi(R)\},
\]
is a singleton; \ie, that there is a unique solution to the nonlinear system given in~\eqref{eq:nonlinear-system}. We can interpret
the set $U(\Delta)$ as the set of reserves for which a given trade $\Delta$ is feasible. Note that any solution
to~\eqref{eq:nonlinear-system}, for given $\Delta$ and $c$, is, by definition, going to be in the intersection of
$U(\Delta) \cap Q(c)$. Because the true reserves $R$ satisfy both equations, it is clear that $U(\Delta) \cap Q(c)$
is nonempty; our goal now is to show that the intersection contains exactly one element, $R$, and therefore
that Eve can correctly recover the reserves by finding a solution to~\eqref{eq:nonlinear-system}.

To show that this intersection is a singleton, it suffices to show that
\[
\psi(kR + \Delta) = \psi(kR)
\]
has the unique solution $k = 1$, because $R \in Q(c)$ and therefore every element of $Q(c)$ is
of the form $kR$ by the previous argument. We will divide this problem into the cases where $k > 1$ and $k < 1$. First, 
assume that $k > 1$, then, by definition of $R$,
\[
\psi(R + \Delta) = \psi(R),
\]
but since $\psi$ is strictly concave, then, for any $0 < \eta < 1$:
\[
\psi(R + \eta\Delta) = \psi(\eta(R + \Delta) + (1-\eta)R) > \eta\psi(R + \Delta) + (1-\eta)\psi(R) = \psi(R).
\]
Setting $\eta = 1/k$ we have that $0 < \eta < 1$, so
\[
\psi(R + (1/k)\Delta) > \psi(R),
\]
or, multiplying on both sides by $k$ and using the homogeneity of $\psi$:
\[
\psi(kR + \Delta) > \psi(kR),
\]
so $k > 1$ cannot be a solution. To show the $0 < k < 1$ case, we will show the contrapositive: if $0 < k < 1$ and
\[
\psi(kR + \Delta) \ge \psi(kR),
\]
then $\psi(R + \Delta) \ne \psi(R)$. This follows from a nearly identical proof as the above: we have that
\[
\psi(kR + k\Delta) > k\psi(kR + \Delta) + (1-k)\psi(kR) \ge  \psi(kR),
\]
where the first inequality follows from the strict concavity of $\psi$, while the second follows by assumption. Then we
immediately have:
\[
\psi(R + \Delta) = \frac1k \psi(kR + k\Delta) > \frac1k \psi(kR) = \psi(R),
\]
so $\psi(R + \Delta) \ne \psi(R)$ as required. The contrapositive then implies that, if $R$ is a solution,
we must have that
\[
\psi(kR + \Delta) < \psi(kR).
\]
for $0 < k < 1$. Combining both statements gives that
\[
\psi(kR + \Delta) = \psi(kR)
\]
if, and only if, $k=1$.

\paragraph{Discussion.} The proof essentially makes use of two important `tricks,' which
might be generalizable to the case where the function $\psi$ is not homogeneous of any 
nonzero degree.

The first is that, because the $\alpha$-superlevel set of $\psi$ is
strictly convex, then any supporting hyperplane maps to a unique reserve value $R$
(depending, implicitly, on $\alpha$). In some sense, this provides a way of `identifying'
reserves, at some fixed, but potentially unknown, liquidity, with a marginal price $c$.
In our case, we used the homogeneity to prove an explicit form for the set $Q(c)$,
but this is likely unnecessary and $Q(c)$ likely satisfies some similarly useful property
without requiring the homogeneity of $\psi$. (We note that Curve is a counterexample to
the plausible conjecture that $Q(c)$ is always a ray for all strictly convex trading functions,
since scaling the reserves of Curve by any nonzero constant will change the marginal price with respect
to any num\'eraire whenever $\alpha, \beta > 0$; \cf,~\cite[\S2.4]{angerisImprovedPriceOracles2020}.)

The
second is that, by the monotonicity and strict concavity of $\psi$, a given trade $\Delta$
feasible for some reserves $R$ with marginal price $c$ will either be too expensive when the CFMM has more liquidity
than $R$ (\ie, there is a strictly better trade $\Delta'$ for the trader that is feasible)
or infeasible when there is less liquidity than $R$. In the case where $\psi$ is homogeneous,
`more' and `less' liquidity, at fixed marginal price $c$ is very easy to identify, since
all possible reserves lie along a ray and are totally ordered, but a more general construction will require some care.

\section{Extensions}\label{sec:extensions}
There are a number of basic extensions which are available to this attack and for which the proof
still holds with either slight or no modifications.

\paragraph{Nonzero fees.} In the case that the function has nonzero fees, \ie, if the CFMM must instead
satisfy
\[
\psi(R + \gamma\Delta_+ - \Delta_-) = \psi(R),
\]
where $\Delta_+$ is the vector whose nonzero entries are the nonnegative entries of $\Delta$ (with all
other entries equal to zero) and similarly for $\Delta_-$, except with the nonpositive entries of $\Delta$. In
this case, the `feasibility' condition is changed slightly for a trade, but the proof of uniqueness remains
otherwise identical.

\paragraph{General homogeneity.} Although we assume 1-homogeneity for a slightly cleaner exposition,
the proof is nearly identical if 1-homogeneity of $\psi$ is replaced with $p$-homogeneity of $\psi$;
\ie, for $\lambda \ge 0$
\[
\psi(\lambda R) = \lambda^p \psi(R),
\]
with $p \ne 0$. The case of $p = 0$ is unlikely to be useful since it would imply that the CFMM
is not sensitive with respect to scaling of the reserves (\ie, it is liquidity-insensitive) which
would imply that liquidity provision does not change the dynamics of the CFMM. We expect similar
problems in liquidity-insensitive CFMMs as those of the classical
AMMs~\cite{othmanLiquiditySensitiveAutomatedMarket2011}.

\paragraph{Unknown marginal price.} In the case that the marginal price is unknown; \eg, it cannot
be accessed directly, it is not difficult for Eve to compute an arbitrarily-good approximation by
performing $n$ queries. In particular, let $\Delta^i \in \reals^n$ for $i=1, \dots, n$ be any
$n$ feasible trades, then we know that, by the concavity of $\psi$:
\[
\psi(R+\Delta^i) < \psi(R) + \lambda c^T\Delta^i, \qquad i=1, \dots, 2n,
\]
where $\lambda c = \nabla \psi(R)$ with $\lambda > 0$ fixed. But, since $\Delta^i$ is a feasible
trade, we have that $\psi(R) = \psi(R + \Delta^i)$, so $c$ must satisfy
\[
\begin{aligned}
	c^T\Delta^i &\ge \eps\ones, \quad i=1, \dots, n\\
	c &\ge 0,
\end{aligned}
\]
where $\ones$ is the all-ones vector and $\eps > 0$ satisfies
\[
\eps \le \min_i (\psi(R) - \psi(R+\Delta^i)).
\]
We note that this is not essential, since the equations can be scaled by $1/\eps$
to recover $c$ up to a constant multiple, so it suffices to solve the following
system of inequalities:
\begin{equation}\label{eq:marginal-price}
\begin{aligned}
	c^T\Delta^i &\ge \ones, \quad i=1, \dots, n\\
	c &\ge 0.
\end{aligned}
\end{equation}
This sets up a system of $2n$ inequalities in $n$ variables, which has a unique solution, up to a nonnegative constant multiple,
provided some basic
conditions on the trades $\Delta^i$. In particular, uniqueness of a solution can be shown since
$\psi$ is strictly concave, under some mild conditions on the queries. The attack then proceeds
as previously stated, replacing querying the marginal price with querying $n$ feasible trades and computing
the marginal price by solving the system of inequalities~\eqref{eq:marginal-price}.

\paragraph{Non-strict concavity.} The more general case where $\psi$ is not strictly concave is rather
more difficult because it implies that there exists some reserve quantities which may map
to the same marginal price and the uniqueness proof need not hold. In some special cases, such as constant sum market makers (\eg, 
mStable), there are simple mitigating strategies:
since the marginal price is fixed everywhere, Eve can make proportionally larger trades until a trade becomes infeasible.
She can then simply perform a binary search to get $\eps$-close to the true reserve quantities in $\log(1/\eps)$ queries.

More general CFMMs whose concave functions are not strictly concave are slightly more difficult, but follow by a similar argument.
In that case, given a fixed marginal price $c$, the set of reserves consistent with $c$, defined as $Q(c)$ above, is a set whose affine 
hull might have dimension greater than 1, but is a convex set. Because these reserves all imply the same marginal price, Eve can 
similarly query
progressively larger trades at the fixed price, until these trades become infeasible. Using the fact that a point in a convex
set can be identified with $n+1$ distances to known points on the set (under mild conditions), Eve can then identify a unique reserve 
$R \in Q(c)$ satisfying all of the above constraints.

Making the uniqueness argument rigorous is rather more involved, even in the case where
$\psi$ is 1-homogeneous, and we do not make it here as most CFMMs are strictly concave in practice.

\section{Mitigating strategies}\label{sec:mitigations}
There are a number of useful conditions for which the proof fails and such conditions might lead
to modifications of CFMMs in order to guarantee privacy. We discuss some basic examples of what
these modifications might look like, but do not prove (nor guarantee) that these modifications
are sufficient for privacy.

\paragraph{Randomness in price.} One immediate example is that Eve requires knowledge of
$c$ in order to reconstruct the reserves. Here, it is possible that the CFMM could add
some amount of randomness (in a similar vein to differential privacy; see, \eg,~\cite{barber2014privacy, dwork2014algorithmic})
to the price in order to prevent Eve from correctly reconstructing the reserve values. Note that
this randomness must be fixed at each block since it would otherwise be easy to approximate the true
price by averaging multiple queries. Additionally, we note that this construction can quickly become
complicated since the price, with randomness, needs to be consistent with the
feasible trades and cannot `leak' too much information after the trades are completed. There is
also the more general problem that such differences in price, due to the added randomness, might
be exploited by arbitrageurs, causing additional losses for the liquidity providers.

\paragraph{Batching orders.} Another possibility is that the CFMM automatically \emph{batches} orders;
\ie, the CFMM waits until several trades $\Delta$ are accepted and updates its reserves only
after all trades have been executed, taking the trades and paying the output all at once.
This leads to two potential difficulties. The first is that the CFMM
needs to ensure that all of the orders, excluding Alice's, are not under Eve's control. (This
could be guaranteed, \eg, if there is always enough trading volume that could make this
attack prohibitively expensive for Eve.) The second is that the CFMM needs to wait until a batch
of orders is specified before executing the trades. This delay can lead to bad user-facing performance
and may threaten the solvency of the system in extreme cases with large price fluctuations.

\paragraph{Further thoughts.} In general, there will likely always be a tradeoff between the
cost users (\ie, liquidity providers or traders) are willing to pay in order to achieve privacy. Such
a cost can be a direct cost, such as higher prices for a fixed trade, or an indirect cost, such as higher risk of failure, resulting
in higher price volatility within such systems. At the moment, it is not clear what price a given
proportion of users are willing to pay in order to ensure privacy. This leads to further question
on where we should allocate the (finite) developer time available for building such systems
in order to ensure the best user experience, while guaranteeing that the systems are safe
in useful ways.

\section{Conclusion}

We have shown that privacy in CFMMs requires more than simply obscuring reserve and trade quantities from possible attackers. In particular,
an attacker can always recover the reserve quantities of a CFMM
given only information that is required for agents to be able to interact with
the system in a meaningful way: \ie, given some amount of coin Alice adds to the CFMM, how much
output can she receive? This attack highlights the difficulty of achieving privacy in DeFi
even under relatively weak adversarial models. We also discussed possible ways
of preventing attackers from knowing the traded quantities, but note that these
are either difficult to achieve in practice or suffer from a degraded user experience.
In general, we suspect that there exist reasonable variations of CFMMs which are privacy-preserving, but their implementation is likely to be neither obvious nor frictionless for the end user.

This note leaves open two major research questions, in order of increasing importance.
First, does the attack outlined here work for all strictly convex CFMMs, not just
those that are 1-homogeneous? We suspect this is the case, but have not been
able to give a reasonable proof except in other special cases. And, second,
what does a privacy-preserving CFMM look like, and
what privacy guarantees can it make? This question is likely much harder,
but also far more important. We suspect that any reasonable progress made
towards answering it is likely to be very useful in both the theory and practice.

\section*{Acknowledgements}
The authors would like to thank Assimakis Kattis for useful discussions regarding
mitigating strategies and Anna Rose for inspiring this note.

\bibliographystyle{alpha}
\bibliography{citations.bib}

\end{document}